\documentclass[12pt]{article}
\usepackage{amssymb}
\newcommand{\fn}{\footnote}
\newtheorem{proposition}{Proposition}[section]  
\newcommand{\bprop}{\medskip\begin{proposition} ~~\\ \it}
\newcommand{\eprop}{\end{proposition} \hfill $\Box$ }
\newtheorem{naming}{Definition}[section]   
\newcommand{\bdefi}{\medskip\begin{naming} ~~\\ \it}
\newcommand{\edefi}{\end{naming} \hfill $\Diamond$ }
\newtheorem{example}{Example}[section]   
\def\bexam{\medskip\begin{example} ~~\\ \rm}
\def\eexam{\end{example} \hfill $\triangle$ }
\newcommand{\sect}[1]{\setcounter{equation}{0}\section{#1}}
\newcommand{\subsect}[1]{\subsection{#1}}

\newcommand{\be}{\begin{equation}}
\newcommand{\ee}{\end{equation}}
\newcommand{\bea}{\begin{eqnarray}}
\newcommand{\eea}{\end{eqnarray}}
\newcommand{\bean}{\begin{eqnarray*}}
\newcommand{\eean}{\end{eqnarray*}}
\newcommand{\nn}{\nonumber}
\newcommand\IC{{\mathbb C}}

\newcommand\II{{\mathbb I}}

\newcommand\IM{{\mathbb M}}
\newcommand\IN{{\mathbb N}}

\newcommand\IZ{{\mathbb Z}}
\newcommand\IR{{\mathbb R}}

\def\abs#1{{\vert#1\vert}}

\def\inf1{{\cal L}^{(1,\infty)}}

\def\bar#1{\overline{#1}}

\def\bra#1{\left\langle #1\right|}
\def\ket#1{\left| #1\right\rangle}
\def\hs#1#2{\left\langle #1|#2\right\rangle}

\def\ca{{\cal A}}
\def\cb{{\cal B}}

\def\raw{\rightarrow}

\def\lrw{\leftrightarrow}
\begin{document}
\setcounter{page}{0}
\thispagestyle{empty}
\begin{flushright}
Trieste-DSM-QM454 \\
May, 1999
\end{flushright}
\vspace{.5cm}
\begin{center}{\Large \bf Projective Modules of Finite Type \\ ~ \\ 
and Monopoles over $S^2$}
\end{center} 
\vspace{1cm}
\centerline{\large  Giovanni Landi}
\vspace{5mm}
\begin{center}
{\it Dipartimento di Scienze Matematiche,
Universit\`a di Trieste \\ P.le Europa 1, I-34127, Trieste, Italy 
\\ and INFN, Sezione di Napoli, Napoli, Italy. \\
landi@mathsun1.univ.trieste.it}
\end{center}
\vspace{2.5cm}
\begin{abstract}
We give a unifying description of all inequivalent vector bundles over the
$2$- dimensional sphere $S^2$ by constructing suitable global projectors $p$ via
equivariant maps. Each projector determines the projective module of finite type of
sections of the corresponding complex rank $1$ vector bundle over $S^2$. 
The canonical connection  $\nabla = p \circ d$ is used to compute the topological
charges. Transposed projectors gives opposite values for the charges, thus showing
that transposition of projectors, although an isomorphism in $K$-theory, is not the
identity map. Also, we construct the partial isometry yielding the equivalence between
the tangent projector (which is trivial in $K$-theory) and the real form of the
charge $2$ projector.
\end{abstract}

\vfill
{\hfill \it This work is dedicated to Jacopo}

\newpage
\sect{Preliminaries and Introduction}
Since the creation of noncommutative geometry \cite{Co,D-V} finite (i.e. of
finite type) projective modules as substitutes for vector bundles are increasingly
being used among (mathematical)-physicists. This substitution is
based on the Serre-Swan's theorem \cite{Sw,Co1} which construct a complete equivalence
between the category of (smooth) vector bundles over a (smooth) compact manifold $M$
and bundle maps, and the category of finite projective modules over the commutative
algebra $C(M)$ of (smooth) functions over $M$ and module morphisms. The space
$\Gamma(M,E)$ of smooth sections of a vector bundle $E \raw M$ over a compact
manifold $M$ is a finite projective module over the commutative algebra $C(M)$ and
every finite projective $C(M)$-module can be realized as the module of sections of
some vector bundle over $M$. 
The correspondence was already used in \cite{Ko} to give an algebraic version of
classical geometry, notably of the notions of connection and covariant derivative.
But it has been with the advent of noncommutative geometry that the
equivalence has received a new emphasis  and has been used, among several other
things, to generalize the concept of vector bundles to noncommutative geometry and
to construct noncommutative gauge and gravity theories. 

In this paper we present a finite-projective-module description of all monopoles
configurations on the $2$-dimensional sphere $S^2$. This will be done by
constructing a suitable global projector $p\in \IM_{\abs{n} + 1}(C(S^2))$, $n\in\IZ$
being the value of the topological charge, which determines the module of sections of
the vector bundles on which monopoles live, as  the image of $p$ in the trivial module
$C(S^2)^{\abs{n} + 1}$ (corresponding to the trivial rank $(\abs{n} + 1)$-vector
bundle over $S^2$).

Now, a local expression for projectors corresponding to monopoles was given in
\cite{mignaco}. Our presentation is a global one which does not use any local chart or
partition of unity. The price we pay for this is that the projector carrying charge
$n$ is a matrix of dimension $(\abs{n} + 1)\times(\abs{n} + 1)$  while in
\cite{mignaco} the projectors were always $2 \times 2$ matrices. Furthermore, our
construction is based on a unifying description in terms of global equivariant maps.
We express the projectors in terms of a more fundamental object, a vector-valued
function of basic equivariant maps. In a sense, we may say that we `{\it deconstruct}'
the projectors \cite{La}. 

The present construction will be generalized to supergeometry in \cite{La1} where we
will report on a construction of `graded monopoles' on the supersphere $S^{2,2}$.
 
A friendly approach to modules of several kind (including finite projective) is in
\cite{book}. In the following we shall avoid writing explicitly the
exterior product symbol for forms.  

\sect{The General Construction}
Let us start by briefly describing the general scheme that will be given in details in
the next Sections. Let $\pi : S^3 \raw S^2$ be the Hopf principal fibration over the
sphere $S^2$ with $U(1)$ as structure group. We shall denote with
$\cb_{\IC} =: C^\infty(S^3,\IC)$ the algebra of $\IC$-valued smooth functions on the
total space $S^3$ while $\ca_{\IC} =: C^\infty(S^2,\IC)$ will be the algebra of
$\IC$-valued smooth functions on the base space $S^2$. The algebra $\ca_{\IC}$ will
not be distinguished from its image in the algebra $\cb_{\IC}$ via pullback.

On $\IC$ there are left actions of the group $U(1)$ and they are labeled by an integer
$n\in\IZ$, two representations corresponding to different integers being inequivalent.
Let
$C^\infty_{(n)}(S^3, \IC)$ denote the collection of corresponding equivariant maps:
\be
\varphi : S^3 ~\raw~ \IC~, ~~\varphi(p\cdot w) = w^{-n}\cdot \varphi(p)~, 
\ee
with $\varphi \in C^\infty_{(n)}(S^3, \IC)$ and for any $p\in S^3$, $w\in U(1)$. The
space $C^\infty_{(n)}(S^3, \IC)$  is a right module over the (pull-back of the) algebra
$\ca_{\IC}$. Moreover, it is well known (see for instance
\cite{Tr2}) that there is a module isomorphism between $C^\infty_{(n)}(S^3, \IC)$ and
the (right) $\ca_{\IC}$-module of sections $\Gamma^\infty(S^3, E^{(n)})$ of the
associated vector bundle $E^{(n)}=S^3\times_{U(1)}\IC$ over $S^2$. In the spirit of
Serre-Swan's theorem \cite{Sw}, the module $\Gamma^\infty(S^3, E^{(n)})$ will be
identified with the image in the trivial, rank $N$, module $(\ca_{\IC})^N$ of a
projector $p\in \IM_N(\ca_{\IC})$, the latter being the algebra of $N\times N$ matrices
with entries in $\ca_{\IC}$, i.e. $\Gamma^\infty(S^3, E^{(n)}) = p (\ca_{\IC})^N$. The
integer $N$ will turn out to be given by 
\be
N = \abs{n} + 1.
\ee
The bundle and the associated projector being of rank $1$ (over $\IC$), the projector
will be written as a ket-bra valued function,
\be\label{urpro}
p=\ket{\psi}\bra{\psi}~, 
\ee
with 
\be\label{urket}
\bra{\psi}=
\left(\psi_1, \dots, \psi_{N} \right)~,
\ee
a specific vector-valued function on $S^3$, thus a specific element of
$(\cb_{\IC})^{N}$, the components being functions $\psi_j \in \cb_{\IC}~, ~j
=1, \dots, N$. The vector-valued function (\ref{urket}) will be normalized,
\be
\hs{\psi}{\psi}=1~,
\ee 
a fact implying that $p$ is a projector
\be
p^2 = \ket{\psi}\hs{\psi}{\psi}\bra{\psi} = p~, ~~~p^\dagger = p~,
\ee
with the symbol ~$^\dagger$~ denoting the adjoint. Furthermore, the normalization will
also imply that $p$ is of rank $1$ over $\IC$ because
\be\label{tragen}
tr(p) = \hs{\psi}{\psi} = 1~.
\ee
In fact, the right end side of (\ref{tragen}) is not the number $1$ but rather the
constant function $1$. Then a normalized integration yields the number $1$
as the value for the rank of the projector and of the associated vector bundle. 

The transformation rule of the vector-valued function $\ket{\psi}$ under the right
action of an element $w \in U(1)$ will be such that the projector $p$ is invariant.
Thus, its entries are functions on the base space $S^2$, that is they are
elements of the algebra $\ca_{\IC}$ and $p \in \IM_{N}(\ca_{\IC})$, 
as it should be.
Elements of $(\ca_{\IC})^{N}$ will be denoted by the symbol 
\be
\ket{\ket{f}}=
\left(
\begin{array}{c}
f_1 \\ \vdots \\ f_{N} 
\end{array}
\right)~,
\ee
with $f_1, \dots, f_{N}$, elements of $\ca_{\IC}$. Then, the module isomorphism
between sections and equivariant maps will be explicitly  given by,
\bea\label{isoseem}
&& \Gamma^\infty(S^2,E^{(n)}) ~\lrw~ C^\infty_{(n)}(S^3, \IC)~, \nn \\
&& \sigma = p\ket{\ket{f}} ~\lrw~ \varphi^\sigma =: \hs{\psi}{\sigma} = 
\hs{\psi}{\ket{f}} = \sum_{j=1}^{N} \psi_j f_j~,
\eea
where we have used the explicit identification $\Gamma^\infty(S^2, E^{(n)}) = p
(\ca_{\IC})^{N}$, $N = \abs{n} + 1$. 

Having the projector, we can define a canonical
connection (the Grassmann connection) on the module of sections by,
\bea\label{con}
&&\nabla =: p \circ d ~:~ \Gamma^\infty(S^2,E^{(n)}) ~\raw~ \Gamma^\infty(S^2,E^{(n)})
\otimes_{\ca_{\IC}} \Omega^1(S^2, \IC)~, \nn \\
&&\nabla\sigma =: \nabla\Big(p\ket{\ket{f}}\Big) = p\Big(\ket{\ket{df}} +
dp\ket{\ket{f}}\Big)~.
\eea
Its curvature $\nabla^2 : \Gamma^\infty(S^2,E^{(n)}) \raw
\Gamma^\infty(S^2,E^{(n)}) \otimes_{\ca_{\IC}} \Omega^2(S^2, \IC)$ 
is found to be
\be\label{cur}
\nabla^2 = p (dp)^2~. 
\ee
By means of a matrix trace, the first Chern class of the vector bundle is given as
\cite{Co},
\be\label{checla}
C_1(p) =: - {1 \over 2 \pi i} ~tr(\nabla^2) 
= - {1 \over 2 \pi i} ~tr (p (dp)^2)~.
\ee
When integrated over $S^2$ it will yield the corresponding Chern number,
\be\label{chenum}
c_1(p) = \int_{S^2} C_1(p) ~.
\ee 
For rank $1$ projectors of the form (\ref{urpro}), the curvature is easily found
to be given by
\be\label{cur1}
\nabla^2 = p (dp)^2 = \ket{\psi}\hs{d \psi}{d \psi} \bra{\psi}~, 
\ee
and the associated Chern form and Chern number are then, 
\be\label{chern}
C_1(p) = - {1 \over 2 \pi i} \hs{d \psi}{d \psi}~, ~~~~
c_1(p) = - {1 \over 2 \pi i} \int_{S^2} \hs{d \psi}{d \psi}~.
\ee

\bigskip
By taking the transpose 
\fn{Since we are considering only self-adjoint idempotents, i.e.
projectors, the transpose is the same as the complex conjugate.} 
of the projector (\ref{urpro}) we still get a projector,
\be
q =: p^{t} = \ket{\phi} \bra{\phi}~,
\ee
with the transposed bra-valued functions given by,
\be\label{bratra}
\bra{\phi}=: \Big(\ket{\psi}\Big)^{t} = \bra{\bar{\psi}} = 
\left(
\bar{\psi}_1, \dots, \bar{\psi}_{N} \right)~. 
\ee
That $q$ is a projector ($q^2=q=q^\dagger$) of rank $1$ (over $\IC$) are
both consequences
of the normalization $ \hs{\phi}{\phi} = \hs{\psi}{\psi} = 1$. But it turns out
that the transposed projector is {\it not} equivalent to the starting one, the
corresponding topological charges differing in sign, i.e.
\be
c_1(p^t) = -c_1(p)~.
\ee
As we shall see, the change in sign comes from the antisymmetry of the exterior product
for forms. Thus, transposing of projectors yields an isomorphism in $K$-theory which is
not the identity map.

Given the connection (\ref{con}), the corresponding connection $1$-form on the
equivariant maps, $A_\nabla \in End_{\cb_{\IC}}(C^\infty_{U(1)}(S^3,\IC))
\otimes_{\cb_{\IC}} \Omega^1(S^3, \IC)$, has a very simple expression in terms of 
the vector-valued function $\ket{\psi}$ \cite{La}, being just
\be\label{confor}
A_\nabla = \hs{\psi}{d \psi}~.
\ee
The associated covariant derivative is given by
\be
\nabla \varphi^\sigma =: \hs{\psi}{\nabla \sigma} 
= \Big(d + \hs{\psi}{d \psi}\Big) \varphi^\sigma~,
\ee 
for any $\sigma \in \Gamma^\infty(S^2,E^{(n)})$ and where we have used the isomorphism
(\ref{isoseem}). The connection form (\ref{confor}) is anti-hermitian, a consequence of
the normalization
$\hs{\psi}{\psi}=1$:
\be
(A_\nabla)^\dagger =: \hs{d\psi}{\psi} = - \hs{\psi}{d \psi} = -A_\nabla~.
\ee

Finally, on the ket-valued function $\ket{\psi}$ there will also be a global {\it
left} action of the unitary group $SU(N) = \{s ~|~ s s^\dagger = 1 \}$ which
preserves the normalization, 
\be
\ket{\psi} ~\mapsto~ \ket{\psi^s} = s \ket{\psi}~, 
~~~ \hs{\psi^s}{\psi^s} = 1~.
\ee
The corresponding transformed projector 
\be
p^s = \ket{\psi^s} \bra{\psi^s} = s \ket{\psi} \bra{\psi} s^\dagger = s p
s^\dagger~,
\ee
is clearly equivalent to the starting one, the partial isometry being $v = s p$;
indeed, one finds $vv^\dagger = p^s$ and $v^\dagger v = p$.  Furthermore, the
connection
$1$-form is left invariant,
\be\label{invcon}
A_{\nabla^s} = \hs{\psi^s}{d \psi^s} = \bra{\psi}s^\dagger s\ket{d \psi} =
A_\nabla~.
\ee
To obtain new (in general gauge non-equivalent) connections one should act with
group elements which do not preserve the normalization. Thus, let $g \in
GL(N; \IC)$
act on the ket-valued function $\ket{\psi}$ by 
\be
\ket{\psi} ~\mapsto~ \ket{\psi^g} = 
\Big[\bra{\psi}g^\dagger g\ket{\psi}\Big]^{-{1 \over 2}} g \ket{\psi} ~.
\ee
The corresponding transformed projector
\bea
p^g &=& \ket{\psi^g} \bra{\psi^g} = 
\bra{\psi}g^\dagger g\ket{\psi}^{-1} g \ket{\psi} \bra{\psi} g^\dagger~,
\nn \\  
&=& \bra{\psi}g^\dagger g\ket{\psi}^{-1} g p g^\dagger
\eea
is again equivalent to the starting one, the partial isometry being now,
\be
v = \Big[\bra{\psi}g^\dagger g\ket{\psi}\Big]^{-{1 \over 2}} ~g p~.
\ee
Indeed,
\bea
&& vv^\dagger =
\bra{\psi}g^\dagger g\ket{\psi}^{-1} ~g p g^\dagger = p^g ~, \nn \\
&& v^\dagger v = \bra{\psi}g^\dagger g\ket{\psi}^{-1} p g^\dagger g p = 
\bra{\psi}g^\dagger g\ket{\psi}^{-1} \ket{\psi}\bra{\psi}
g^\dagger g \ket{\psi}\bra{\psi} = p~.
\eea
The associated connection $1$-form is readily found to be
\bea\label{tracon}
A_{\nabla^g} &=:& \hs{\psi^g}{d \psi^g} \nn \\
&=& {1 \over 2} \bra{\psi}g^\dagger g\ket{\psi}^{-1} 
[\bra{\psi}g^\dagger g\ket{d \psi} - \bra{d \psi}g^\dagger g\ket{\psi}]~.
\eea
Thus, if $g \in SU(N)$, we get back the previous invariance of connections
(\ref{invcon}), while for  $g \in GL(N)$ modulo $SU(N)$ we get new, gauge
non-equivalent connections on the complex line bundle over $S^2$ determined by the
projector $p^g$, line bundle which is (stable) isomorphic to the one determined by the
projector $p$.
 
\sect{The Hopf Fibration over $S^2$}
The $U(1)$ principal fibration $\pi : S^3 \raw S^2$ over the two
dimensional sphere is explicitly realized as follows.
The total space is the three dimensional sphere
\be
S^3 = \{(z_0,z_1) \in \IC^2 ~,~ |z_0|^2 + |z_1|^2 = 1\}~, 
\ee
with right $U(1)$-action 
\be\label{u1act}
S^3 \times U(1) ~\raw~ S^3~, ~~~(z_0,z_1) \cdot w = (z_0 w, z_1 w)~.
\ee
Clearly $|z_0 w|^2 + |z_1 w|^2 = |z_0|^2 + |z_1|^2 = 1$. 
The bundle projection $\pi : S^3 \raw S^2$ is just the Hopf projection and it is 
given by $\pi(z_0,z_1) =: (x_1, x_2, x_3)$, 
\bea\label{s2coord}
&& x_1 = z_0\bar{z}_1 + z_1\bar{z}_0 ~,\nn \\ 
&& x_2 = i(z_0\bar{z}_1 - z_1\bar{z}_0) ~, \nn \\ 
&& x_3 = |z_0|^2 - |z_1|^2 = -1 +2|z_0|^2 = 1 - 2|z_1|^2 ~, 
\eea
and one checks that $\sum_{\mu=1}^3 (x_\mu)^2 = (|z_0|^2 + |z_1|^2)^2 = 1$~. 
The inversion of (\ref{s2coord}) gives the basic ($\IC$-valued) invariant functions
on $S^3$,
\be
|z_0|^2 = {1 \over 2} (1+x_3)~, 
~~~~|z_1|^2 = {1 \over 2} (1-x_3)~,
~~~~z_0 \bar{z}_1 = {1 \over 2} (x_1 -i x_2)~,
\ee 
a generic invariant (polynomial) function on $S^3$ being any function of the
previous variables. Later on we shall also need the volume form of $S^2$ which turns
out to be
\be\label{vols2}
d (vol(S^2)) = x_1 dx_2 dx_3 + x_2 dx_3 dx_1 + x_3 dx_1 dx_2 
= 2 i (dz_0 d\bar{z}_0 + dz_1 d\bar{z}_1)~.
\ee

\subsect{The Equivariant Maps}

Irreducible representations of the group $U(1)$ are labeled by an integer
$n\in\IZ$, any two representations associated with different integers being
inequivalent. They can be explicitly given as left  representations on $\IC$, 
\be\label{repn}
\rho_n ~:~ U(1) \times \IC ~\raw~ \IC~, 
~~~(w,c) \mapsto \rho_n(w) \cdot c =: w^n c~.
\ee
In order to construct the corresponding equivariant maps $\varphi : S^3 \raw \IC$
we shall distinguish between the two cases for which the integer $n$ is negative or
positive. In fact, from now on, we shall take the integer $n$ to be always positive 
and consider the two cases corresponding to $\mp n$.

\bigskip
\noindent
{\bf The equivariant maps.}~ Case $1.$~~~ $-n~, ~~n \in \IN$.

The generic equivariant map $\varphi_{-n} : S^3 \raw \IC$ is of the form
\be\label{equi-n}
\varphi_{-n}(z_0,z_1) = \sum_{k=0}^n(z_0)^{n-k}(z_1)^{k}~ g_k(z_0,z_1)~,
\ee
with $g_k~, ~k =0,1, \dots, n~$, generic $\IC$-valued functions on $S^3$ which are
invariant under the right action of $U(1)$. Indeed,
\bea\label{equi-n1}
\varphi_{-n}((z_0,z_1)w) &=&
\sum_{k=0}^n(z_0 w)^{n-k}(z_1 w)^{k}~ g_k(z_0 w,z_1 w) 
= w^n \sum_{k=0}^n(z_0)^{n-k}(z_1)^{k}~ g_k(z_0,z_1) \nn \\ 
&=& \rho_{-n}(w)^{-1} \cdot \varphi_{-n}(z_0,z_1)~.
\eea
We shall think of the functions $g_k$'s as $\IC$-valued functions on the base space
$S^2$, namely as elements of the algebra $\ca_{\IC}$. The 
collection $C^\infty_{(-n)}(S^3,
\IC)$ of equivariant maps is a right module over the (pull-back of) functions
$\ca_{\IC}$.

\bigskip
\noindent
{\bf The equivariant maps.}~ Case $2.$~~~ $n \in \IN$.

The generic equivariant map $\varphi_n : S^3 \raw \IC$ is of the form
\be\label{equin}
\varphi_n(z_0,z_1) = \sum_{k=0}^n(\bar{z}_0)^{n-k}(\bar{z}_1)^{k}~ f_k(z_0,z_1)~,
\ee
with $f_k~, ~k =0,1, \dots, n~$, generic $\IC$-valued functions on $S^3$ which are
invariant under the right action of $U(1)$. Indeed,
\bea\label{equin1}
\varphi_n((z_0,z_1)w) &=&
\sum_{k=0}^n(\bar{w} \bar{z}_0)^{n-k}(\bar{w} \bar{z}_1)^{k}~ f_k(z_0 w,z_1 w) 
= \bar{w}^n \sum_{k=0}^n(\bar{z}_0)^{n-k}(\bar{z}_1)^{k}~ f_k(z_0,z_1) \nn \\ 
&=& \rho_n(w)^{-1} \cdot \varphi_n(z_0,z_1)~.
\eea
As before, we shall think of the functions $f_k$'s as elements of the algebra
$\ca_{\IC}$. And the collection $C^\infty_{(n)}(S^3, \IC)$ of equivariant
maps will again be
a right module over $\ca_{\IC}$.

\subsect{The Projectors and their Charges}

We are now ready to introduce the projectors. Again we shall take the integer $n$
to be positive and keep separated the two cases corresponding to $\mp n$.

\bigskip
\noindent
{\bf The construction of the projector.}~ Case $1.$~~~ $-n~, ~~n \in \IN$.

Consider the vector-valued function with ($n+1$)-components given by,
\be\label{mon-n}
\bra{\psi_{-n}}=: 
\left((z_0)^n, \dots, 
~\sqrt{ 
{\scriptstyle        
 \addtolength{\arraycolsep}{-.5\arraycolsep}
 \renewcommand{\arraystretch}{0.5}
 \left( 
\begin{array}{l}
 \scriptstyle n \\
 \scriptstyle k  
\end{array} \scriptstyle \right)} 
}
(z_0)^{n-k} (z_1)^k~, \dots, (z_1)^n \right)~,
\ee
where 
\be\left(
\begin{array}{l}
n \nonumber \\ k
\end{array}
\right) = {n! \over{k! (n-k)!}}~, ~~~ k = 0,1, \dots, n~, 
\ee
are the binomial coefficients. The vector-valued function (\ref{mon-n}) is normalized,
\be
\hs{\psi_{-n}}{\psi_{-n}}= \sum_{k=0}^n 
{\scriptstyle        
 \addtolength{\arraycolsep}{-.5\arraycolsep}
 \renewcommand{\arraystretch}{0.5}
 \left( 
\begin{array}{l}
 \scriptstyle n \\
 \scriptstyle k  
\end{array} \scriptstyle \right)} 
(z_0)^{n-k} (z_1)^k (\bar{z}_0)^{n-k} (\bar{z}_1)^k = (\abs{z_0}^2 +
\abs{z_1}^2)^n = 1~.
\ee
Then, we can construct a projector in $\IM_{n+1}(\ca_{\IC})$ by
\be\label{pro-n}
p_{-n} =: \ket{\psi_{-n}} \bra{\psi_{-n}}~.
\ee
It is clear that $p_{-n}$ is a projector,
\bea
&& p_{-n}^2 =: \ket{\psi_{-n}} \hs{\psi_{-n}}{\psi_{-n}} \bra{\psi_{-n}} 
= \ket{\psi_{-n}} \bra{\psi_{-n}} = p_{-n}~, \nn \\
&& p_{-n}^\dagger = p_{-n}~.
\eea
Moreover, it is of rank $1$ because its trace is the constant function
$1$, \be
tr p_{-n} = \hs{\psi_{-n}}{\psi_{-n}} = 1~.
\ee
The $U(1)$-action (\ref{u1act}) will transform the vector (\ref{mon-n})
multiplicatively,
\be
\bra{\psi_{-n}} ~\mapsto~ \bra{(\psi_{-n})^w} = w^n \bra{\psi_{-n}}~, ~~~\forall ~w \in
U(1)~.
\ee
As a consequence the projector $p_{-n}$ is invariant,
\be
p_{-n} ~\mapsto~ (p_{-n})^w = \ket{(\psi_{-n})^w}\bra{(\psi_{-n})^w} =
\ket{\psi_{-n}}\bar{w}^n w^n \bra{\psi_{-n}} = \ket{\psi_{-n}}
\bra{\psi_{-n}} = p_{-n}
\ee 
(being $\bar{w} w = 1$), and its entries are functions on the base space $S^2$, that is
they are elements of
$\ca_{\IC}$ as it should be. Thus, the right module of sections $\Gamma^\infty(S^2,
E^{(-n)})$ of the associated bundle is identified with the image of $p_{-n}$ in the
trivial rank
$n+1$ module $(\ca_{\IC})^{n+1}$ and the module isomorphism between sections and
equivariant maps is given by,
\bea
\Gamma^\infty(S^2,E^{(-n)}) &\lrw& C^\infty_{(-n)}(S^3, \IC)~, \nn \\
\sigma = p_{-n} \ket{\ket{g}} &\lrw& \varphi^\sigma_{-n} = 
\hs{\psi_{-n}}{\sigma} \nn \\
\varphi^\sigma_{-n}(z_0,z_1) &=& \bra{\psi_{-n}} \left(
\begin{array}{c}
g_0 \\ \vdots \\ g_n
\end{array}
\right) 
= \sum_{k=0}^n 
\sqrt{ 
{\scriptstyle        
 \addtolength{\arraycolsep}{-.5\arraycolsep}
 \renewcommand{\arraystretch}{0.5}
 \left( 
\begin{array}{l}
 \scriptstyle n \\
 \scriptstyle k  
\end{array} \scriptstyle \right)} 
}
~(z_0)^{n-k}(z_1)^{k}~ g_k(z_0,z_1) ~, ~~~~~ 
\eea
with $g_0, \dots, g_n$ generic elements in $\ca_{\IC}$. By comparison with
(\ref{equi-n}) it is obvious that the previous map is a module isomorphism, the extra
factors $\sqrt{ 
{\scriptstyle        
 \addtolength{\arraycolsep}{-.5\arraycolsep}
 \renewcommand{\arraystretch}{0.5}
 \left( 
\begin{array}{l}
 \scriptstyle n \\
 \scriptstyle k  
\end{array} \scriptstyle \right)} 
}
$ being inessential to this purpose since they could be absorbed in a
redefinition of the functions.

The canonical connection associated with the projector $p_{-n}$,
\be
\nabla = p_{-n} \circ d ~:~ \Gamma^\infty(S^2,E^{(-n)}) ~\raw~
\Gamma^\infty(S^2,E^{(-n)}) \otimes_{\ca_{\IC}} \Omega^1(S^2, \IC),
\ee
has curvature given by
\be\label{cur-n}
\nabla^2 = p_{-n}(dp_{-n})^2 = \ket{\psi_{-n}}\hs{d \psi_{-n}}{d \psi_{-n}}
\bra{\psi_{-n}}~. 
\ee
The corresponding Chern number is  
\be\label{cn-n}
c_1(p_{-n}) =: - {1 \over 2 \pi i} \int_{S^2}~tr (p_{-n} (dp_{-n})^2) 
= - {1 \over 2 \pi i} \int_{S^2} \hs{d \psi_{-n}}{d \psi_{-n}}~. 
\ee
Now, a lengthy but straightforward computation shows that
\be
\hs{d \psi_{-n}}{d \psi_{-n}} = n (dz_0 d\bar{z}_0 +  dz_1 d\bar{z}_1) 
= {n \over 2 i}  d(vol(S^2))~,
\ee
which, when substituted in (\ref{cn-n}) gives, 
\be\label{cn-n1}
c_1(p_{-n}) = {n \over 4 \pi} \int_{S^2} d (vol(S^2)) = 
{n \over 4 \pi} 4 \pi = n~.
\ee

\bigskip
\noindent
{\bf The construction of the projector.}~ Case $2.$~~~ $n \in \IN$.

Consider the vector-valued function with ($n+1$)-components given by,
\be\label{monn}
\bra{\psi_{n}}=: 
\left((\bar{z}_0)^n, \dots, 
~\sqrt{ 
{\scriptstyle        
 \addtolength{\arraycolsep}{-.5\arraycolsep}
 \renewcommand{\arraystretch}{0.5}
 \left( 
\begin{array}{l}
 \scriptstyle n \\
 \scriptstyle k  
\end{array} \scriptstyle \right)} 
}
(\bar{z}_0)^{n-k} (\bar{z}_1)^k~, \dots, (\bar{z}_1)^n \right)~,
\ee
which is normalized,
\be
\hs{\psi_{n}}{\psi_{n}}= \sum_{k=0}^n 
{\scriptstyle        
 \addtolength{\arraycolsep}{-.5\arraycolsep}
 \renewcommand{\arraystretch}{0.5}
 \left( 
\begin{array}{l}
 \scriptstyle n \\
 \scriptstyle k  
\end{array} \scriptstyle \right)} 
(\bar{z}_0)^{n-k} (\bar{z}_1)^k (z_0)^{n-k} (z_1)^k = (\abs{z_0}^2
+
\abs{z_1}^2)^n = 1~.
\ee
As before, a projector in $\IM_{n+1}(\ca_{\IC})$ is constructed by
\be\label{pron}
p_{n} =: \ket{\psi_{n}} \bra{\psi_{n}}~.
\ee
Indeed $p_{n}^2 =: \ket{\psi_{n}} \hs{\psi_{n}}{\psi_{n}} \bra{\psi_{n}} 
= \ket{\psi_{n}} \bra{\psi_{n}} = p_{n}$ and $p_{n}^\dagger = p_{n}$.
And the projector $p_{n}$ is of rank $1$ because its trace is the constant
function $1$, $tr p_{n} = \hs{\psi_{n}}{\psi_{n}} = 1$.

The $U(1)$-action (\ref{u1act}) will now transform the vector (\ref{monn}) by,
\be
\bra{\psi_{n}} ~\mapsto~ \bra{(\psi_{n})^w} = \bar{w}^n \bra{\psi_{n}}~, ~~~\forall ~w
\in U(1)~.
\ee
As a consequence, the projector $p_{n}$ is invariant,
\be
p_{n} ~\mapsto~ (p_{n})^w = \ket{(\psi_{n})^w}\bra{(\psi_{n})^w} =
\ket{\psi_{n}} w^n\bar{w}^n \bra{\psi_{n}} = \ket{\psi_{n}}
\bra{\psi_{n}} = p_{n}
\ee 
(being $w\bar{w} = 1$),  and its entries are
again functions on the base space $S^2$, that is they are elements of $\ca_{\IC}$.
Thus, also the right module of sections $\Gamma^\infty(S^2, E^{(n)})$ is identified
with the image of $p_{n}$ in the trivial rank $n+1$ module
$(\ca_{\IC})^{n+1}$.  The module isomorphism between sections and equivariant maps
is now given by,
\bea
\Gamma^\infty(S^2,E^{(n)}) &\lrw& C^\infty_{(n)}(S^3, \IC)~, \nn \\
\sigma = p_{n} \ket{\ket{f}} &\lrw& \varphi^\sigma_{n} = 
\hs{\psi_{n}}{\sigma} \nn \\
\varphi^\sigma_{n}(z_0,z_1) &=& 
\bra{\psi_{n}} \left(
\begin{array}{c}
f_0 \\ \vdots \\ f_n
\end{array}
\right)  
= \sum_{k=0}^n \sqrt{ 
{\scriptstyle        
 \addtolength{\arraycolsep}{-.5\arraycolsep}
 \renewcommand{\arraystretch}{0.5}
 \left( 
\begin{array}{l}
 \scriptstyle n \\
 \scriptstyle k  
\end{array} \scriptstyle \right)} 
}
~(\bar{z}_0)^{n-k}(\bar{z}_1)^{k}~ f_k(z_0,z_1) ~,  
\eea
with $f_0, \dots, f_n$ generic elements in $\ca_{\IC}$. By comparison with
(\ref{equin}) it is obvious that the previous map is a module isomorphism
(again the extra
factors could be absorbed in a redefinition of the functions).

The canonical connection associated with the projector $p_{n}$,
\be
\nabla = p_{n} \circ d ~:~ \Gamma^\infty(S^2,E^{(n)}) ~\raw~
\Gamma^\infty(S^2,E^{(n)}) \otimes_{\ca_{\IC}} \Omega^1(S^2, \IC),
\ee
has curvature given by
$ \nabla^2 = p_{n}(dp_{n})^2 = \ket{\psi_{n}}\hs{d \psi_{n}}{d \psi_{n}}
\bra{\psi_{n}}$, and 
the corresponding Chern number is  
\be\label{cnn}
c_1(p_{n}) =: - {1 \over 2 \pi i} \int_{S^2}~tr (p_{n} (dp_{n})^2) 
= - {1 \over 2 \pi i} \int_{S^2} \hs{d \psi_{n}}{d \psi_{n}}~. 
\ee
Now, the vector-valued functions $\bra{\psi_{n}}$ and $\bra{\psi_{-n}}$ transform one
into the other by the exchange  
$z_0 \lrw \bar{z}_0$ and $z_1 \lrw \bar{z}_1$. Thus, as a consequence of the
antisymmetry of the wedge product for $1$-forms one has, 
\be
\hs{d \psi_{n}}{d \psi_{n}} = - \hs{d \psi_{-n}}{d \psi_{-n}} = -n (dz_0 d\bar{z}_0 + 
dz_1 d\bar{z}_1)  = -{n \over 2 i}  d(vol(S^2))~,
\ee
which, when substituted in (\ref{cnn}) gives, 
\be\label{cnn1}
c_1(p_{n}) = -{n \over 4 \pi} \int_{S^2} d (vol(S^2)) = 
-{n \over 4 \pi} 4 \pi = -n~.
\ee

In fact, the functions $\bra{\psi_{n}}$ and $\bra{\psi_{-n}}$ are one the
transposed of the other 
\footnote{As already remarked, in
our case transposition is the same as complex conjugation.}, that is, 
\be\label{bratra-+}
\bra{\psi_{n}} = (\ket{\psi_{-n}})^{t} = \bra{\bar{\psi_{-n}}}~, 
\ee
and the corresponding projectors are related by transposition,
\be
p_{n} = (p_{-n})^{t}~.
\ee
Thus, by transposing a projector we get an inequivalent one \fn{Unless the projector
is the identity.}. This inequivalence is a manifestation of the fact that transposing
of projectors yields an isomorphism in the reduced group $K$-theory
group $\tilde{K}(S^2)$, which is not the identity map.

\bigskip
\noindent {\bf Examples.} Here we give the explicit projectors corresponding to the
lowest values of the charges, $\pm1$, while in the next section we give the ones
corresponding to charge $\pm2$. \\
By using the definition (\ref{s2coord}) for the
coordinate functions on $S^2$, we find that
\bea\label{pro+-} 
&&p_{-1} = 
\left(
\begin{array}{cc}
\abs{z}_0^2 & z_1\bar{z}_0 \\ 
z_0\bar{z}_1 & \abs{z}_1^2
\end{array}
\right) = 
{1 \over 2} \left(
\begin{array}{cc}
1 + x_3 & x_1 + i x_2 \\ 
x_1 - i x_2 & 1 - x_3
\end{array}
\right)~, \nn \\
&& ~\nn \\
&& ~p_{1} =
\left(
\begin{array}{cc}
\abs{z}_0^2 & z_0\bar{z}_1 \\ 
z_1\bar{z}_0 & \abs{z}_1^2
\end{array}
\right) = 
{1 \over 2} \left(
\begin{array}{cc}
1 + x_3 & x_1 - i x_2 \\ 
x_1 + i x_2 & 1 - x_3
\end{array}
\right)~. 
\eea
It is evident that these projectors are one the transposed (or
equivalently, the complex conjugate) of the other.

\subsect{The Monopole Connections}

We are now ready to construct explicitly the monopole connections. The connection
$1$-forms (\ref{confor}) associated with the projectors
$p_{\mp n}$ are given by  
\be\label{confor-+}
A_{\mp n} = \hs{\psi_{\mp n}}{d \psi_{\mp n}}~.
\ee
They are anti-hermitian, 
\be
(A_{\mp n})^\dagger = \hs{\psi_{d \mp n}}{\psi_{\mp n}} 
= - \hs{\psi_{\mp n}}{d \psi_{\mp n}} = - A_{\mp n} ~,
\ee
so they are valued in $i \IR$, the Lie algebra of $U(1)$. A straightforward
computation yields
\be
A_{\mp n} = \mp n (\bar{z}_0 d z_0 + \bar{z}_1 d z_1) = \mp n A_1~, 
\ee
with 
\be
A_1 = \bar{z}_0 d z_0 + \bar{z}_1 d z_1
\ee
the charge $-1$ monopole connection form \cite{Ma,Tr2}. In \cite{Tr1} a local
expression of the connection $A_{-n}$ (corresponding, we recall, to charge $n$) was
given as the pull back to $S^2$ of the Hodge form of the projective space $CP^n$.
There, by thinking of the functions $z_0, z_1$ as homogeneous coordinates on $S^2$, 
the vector-valued function $\bra{\psi_{-n}}$ in (\ref{mon-n}) was used to embed $S^2$
into $CP^n$. 

Finally, the invariance (\ref{invcon}) states the invariance of the
connection $1$-form (\ref{confor-+}) under the global left action of $SU(N)$. Gauge
non-equivalent connections are obtained by the formula (\ref{tracon}),
\be\label{tracon-+}
A_{\mp n}^g = {1 \over 2} 
\bra{\psi_{\mp n}}g^\dagger g\ket{\psi_{\mp n}}^{-1} 
~\Big[\bra{\psi_{\mp n}}g^\dagger g\ket{d \psi_{\mp n}} - \bra{d \psi_{\mp n}}g^\dagger
g\ket{\psi_{\mp n}}\Big]~,
\ee
with $g \in GL(N;\IC)$ modulo $SU(N)$ and $\bra{\psi_{\mp n}}$ given respectively by
(\ref{mon-n}) and (\ref{monn}). 

A description of gauge theories in terms of projectors has been suggested
in \cite{D-Vpro}. To our knowledge, since then there has not
been further work in this direction.

\sect{The Tangent Projector vs the Charge $2$ Projector}

That the bundle $TS^2$ tangent to $S^2$, although not trivial as a bundle, is trivial
in $K$-theory (stable triviality) is well known \cite{Ka} and it is a consequence of
the fact that by adding to $TS^2$ the real rank $1$ trivial bundle one gets the
real rank $3$ trivial bundle.  On the other side, it is also well known \cite{Tr2}
that $TS^2$ can be identified with the real form of the complex charge $2$ monopole
bundle over $S^2$. This identification is an instance of the general result that
equate the top Chern class of a complex vector bundle with the Euler class of the real
form of the bundle. We shall prove this equivalence at level of $K$-theory by
constructing explicitly the partial isometry between the tangent projector and the real
form of the charge $2$ monopole projector. This will also show that classes which are
not trivial in complex $K$-theory may become trivial when translated in real $K$-theory.
 
\subsect{The Tangent Projector}
We shall use real cartesian coordinates $(x_1, x_2, x_3)~, ~\sum_{\mu=1}^3
(x_\mu)^2 = 1$, for the sphere $S^2$ and denote with $\ca_{\IR} =: C^\infty(S^3,\IR)$
the algebra of smooth real-valued functions on $S^2$. The (module of smooth
sections of the) normal bundle over $S^2$ is realized as the image, in the trivial,
real rank $3$ module $(\ca_{\IR})^3$, of the {\it normal projector}
\be\label{pronor}
p_{nor} = \ket{\psi_{nor}}\bra{\psi_{nor}}~, ~~~\bra{\psi_{nor}} = (x_1, x_2,
x_3)~.
\ee
It is clear that $p_{nor}$ is a projector, $p_{nor}^2 = p_{nor} = p_{nor}^\dagger$, of
real rank $1$. Moreover, $p_{nor}$ is stable (and altogether) trivial as it can be
inferred by computing its Chern \fn{In fact, Pontryagin for real bundles.} $1$-form,
\be
C_1(p_{nor}) =: - {1 \over 2 \pi i} tr (p_{nor}(d p_{nor})^2) = - {1 \over 2 \pi i}
\hs{d \psi_{nor}}{d \psi_{nor}} = 0~.
\ee 

Then, the (module of smooth sections of the) tangent bundle is simply realized as the
image in $(\ca_{\IR})^3$ of the {\it tangent projector}
\bea\label{protan}
p_{tan} &=& \II - p_{nor} = \II - \ket{\psi_{nor}}\bra{\psi_{nor}}~, \nn \\
&~& \nn \\
&=& \left(
\begin{array}{ccc} 
1 - (x_1)^2  & -x_1 x_2     & -x_1 x_3 \\
-x_1 x_2     & 1 - (x_2)^2  & -x_2 x_3   \\
-x_1 x_3     & -x_2 x_3     & 1 - (x_3)^2 \\
\end{array}
\right)~.
\eea
That the tangent bundle is of real rank $2$ is translated in the fact that the
trace of $p_{tan}$ is equal to the constant function $2$, $tr(p_{tan}) = 2$. The
tangent  bundle is stable trivial as well since its topological charge vanishes.
Indeed, by using the fact that
\be
\hs{\psi_{nor}}{d \psi_{nor}} = x_1 d x_1 + x_2 d x_2 + x_3 d x_3 = 0~,
\ee
it is straightforward to find, 
\bea
p_{tan}(d p_{tan})^2 &=& \ket{d \psi_{nor}}\bra{d \psi_{nor}} \nn \\
&& \nn \\
&=& dx_1 dx_2 
~{\scriptstyle        
 \addtolength{\arraycolsep}{-.5\arraycolsep}
 \renewcommand{\arraystretch}{0.5}
 \left( \begin{array}{ccc}
 \scriptstyle 0 & \scriptstyle 1 & \scriptstyle 0 \\
 \scriptstyle -1 & \scriptstyle 0 & \scriptstyle 0 \\
 \scriptstyle 0 & \scriptstyle 0 & \scriptstyle 0 
\end{array} \scriptstyle
\right)}
+ dx_2 dx_3 
~{\scriptstyle        
 \addtolength{\arraycolsep}{-.5\arraycolsep}
 \renewcommand{\arraystretch}{0.5}
 \left( \begin{array}{ccc}
 \scriptstyle 0 & \scriptstyle 0 & \scriptstyle 0 \\
 \scriptstyle 0 & \scriptstyle 0 & \scriptstyle 1 \\
 \scriptstyle 0 & \scriptstyle -1 & \scriptstyle 0 
\end{array} \scriptstyle
\right)}
+ dx_3 dx_1 
~{\scriptstyle        
 \addtolength{\arraycolsep}{-.5\arraycolsep}
 \renewcommand{\arraystretch}{0.5}
 \left( \begin{array}{ccc}
 \scriptstyle 0 & \scriptstyle 0 & \scriptstyle -1 \\
 \scriptstyle 0 & \scriptstyle 0 & \scriptstyle 0 \\
 \scriptstyle 1 & \scriptstyle 0 & \scriptstyle 0 
\end{array} \scriptstyle
\right)}~.
\eea 
As a consequence,
\be
C_1(p_{tan}) =: - {1 \over 2 \pi i} tr (p_{tan}(d p_{tan})^2) 
= - {1 \over 2 \pi i} tr \left(\ket{d \psi_{nor}}\bra{d \psi_{nor}} \right) = 0~.
\ee 

For later convenience, we need to express the tangent projector as a sum of three
pieces. Let us then introduce the three vector fields on $S^2$ which generate the
action of $SU(2)$ on $S^2$. They are given by
\be\label{infrot}
V_l = \sum_{j,k=0}^3 \varepsilon_{ljk} ~x_j {\partial \over \partial x_k}~, ~~~l =
1,2,3~.
\ee
and clearly they are not independent; indeed, 
\be
\sum_{l=0}^3 ~x_l V_l = 0~.
\ee
We shall write the vector fields (\ref{infrot}) as vector-valued functions on $S^2$,
\be
\bra{V_1} = (0, -x_3, x_2)~, ~~\bra{V_2} = (x_3, 0, -x_1)~, 
~~\bra{V_3} = (-x_2, x_1, 0)~.
\ee
Then, it is an easy computation to show that the tangent projector can be
written as,
\be\label{protansum}
p_{tan} = \ket{V_1}\bra{V_1} + \ket{V_2}\bra{V_2} + \ket{V_3}\bra{V_3}~.
\ee
Finally, the following properties are easily established
\bea
&& \hs{\psi_{nor}}{V_l} = 0~, ~~~l = 1,2,3 ~. \nn \\
&& p_{tan}\ket{V_l} = \ket{V_l}~, ~~~l = 1,2,3~. \nn \\
&& (p_{tan})_{kl} = \hs{V_k}{V_l}~, ~~~k, l = 1,2,3~. \nn \\
&& tr (p_{tan}) = \sum_{l=0}^3 \hs{V_l}{V_l} = 2 \sum_{\mu=1}^3 (x_\mu)^2 = 2~.
\eea

\subsect{The Charge $2$ Projector and its Real Form}
It turns out that in order to prove the equivalence we are after to, it is best
to `change bases' for the equivariant functions and as a consequence to
construct a charge $2$ projector which is not the same as the one given by
(\ref{pro-n}) but it is of course equivalent to it. 
Let us then consider the following vector-valued equivariant map on $S^3$,
\be\label{tilbra-2}
\bra{\widetilde{\psi}_{-2}} = {1 \over \sqrt{2}} \Big((z_1)^2 - (z_0)^2, ~
(z_1)^2 + (z_0)^2, ~2 z_0 z_1 \Big)
\ee
(we recall that the label $-2$ characterizes the type of representation of $U(1)$
and that it corresponds to charge $2$ as we shall also see presently).
The vector-valued function (\ref{tilbra-2}) is normalized,
\be
\hs{\widetilde{\psi}_{-2}}{\widetilde{\psi}_{-2}} = \left(\abs{z}_0^2 +
\abs{z}_1^2 \right)^2 = 1~. 
\ee 
As a consequence, the following is a complex rank $1$ projector in $\IM_3(\ca_{\IC})$,
\be\label{tilpro-2}
\widetilde{p}_{-2} =: \ket{\widetilde{\psi}_{-2}}\bra{\widetilde{\psi}_{-2}}~.
\ee
That the projector $\widetilde{p}_{-2}$ is equivalent to the projector $p_{-2}$
given by (\ref{pro-n}) for the value $n=2$, is best seen by computing its topological
charge. A simple computation shows that
\be
\hs{d \widetilde{\psi}_{-2}}{d \widetilde{\psi}_{-2}} = 2 (d z_0 d\bar{z}_0 +
d z_1 d\bar{z}_1) = {1 \over i}  d(vol(S^2))~, 
\ee
which, in turns, gives
\be\label{tilcn-n1}
c_1(\widetilde{p}_{-2}) = - {1 \over 2 \pi i} \int_{S^2} \hs{d
\widetilde{\psi}_{-2}}{d \widetilde{\psi}_{-2}} = {1 \over 2 \pi} \int_{S^2} d
(vol(S^2)) = 2~,
\ee
as it should be. This shows the equivalence between $\widetilde{p}_{-2}$ and
$p_{-2}$ (of course one could also directly construct the corresponding partial
isometry).

Next, we express the projector (\ref{tilpro-2}) in terms of the coordinate functions
on $S^2$. It turns out that
\be\label{tilpro-2car}
\widetilde{p}_{-2} = {1 \over 2}
\left(
\begin{array}{ccc}
1 - (x_1)^2  & -x_3 - i x_1 x_2   & -i x_2 - x_1 x_3 \\
& & \\
-x_3 + i x_1 x_2   & 1 - (x_2)^2  & x_1 + i x_2 x_3  \\
& & \\
i x_2 - x_1 x_3   & x_1 - i x_2 x_3    & 1 - (x_3)^2
\end{array}
\right)~.
\ee
From the general considerations described before, the transpose of this projector
would carry charge $-2$.

Let us now turn to real forms.
The real form $(\widetilde{p}_{-2})^{\IR}$ of the projector $\widetilde{p}_{-2}$ in
(\ref{tilpro-2car}) will be a projector in $\IM_6(\ca_{\IR})$ and it is obtained by the
substitution
\be
a + i b ~\mapsto~ 
\left(
\begin{array}{cc}
a & -b \\
b & a 
\end{array}
\right)~, ~~~\forall ~a, b \in \IR~.
\ee
One finds that 
\be\label{tilpro-2carrea}
(\widetilde{p}_{-2})^{\IR} = {1 \over 2}
\left(
\begin{array}{cccccc}
1 - (x_1)^2 & 0 & -x_3 & x_1 x_2 & - x_1 x_3 & x_2  \\
~         \\
0 & 1 - (x_1)^2 & -x_1 x_2 & -x_3 & -x_2 & - x_1 x_3  \\         
~         \\ 
-x_3 & - x_1 x_2 & 1 - (x_2)^2 & 0 & x_1 & -x_2 x_3  \\ 
~         \\
x_1 x_2 & -x_3 & 0 & 1 - (x_2)^2 & x_2 x_3 & x_1  \\ 
~         \\
-x_1 x_3 & -x_2 & x_1 & x_2 x_3 & 1 - (x_3)^2 & 0  \\ 
~         \\
x_2 & -x_1 x_3 & -x_2 x_3 & x_1 & 0 & 1 - (x_3)^2  \\ 
\end{array}
\right)~.
\ee

\noindent
That $(\widetilde{p}_{-2})^{\IR}$ is a projector will also be evident from the
analysis of next Section.

\subsect{The Partial Isometry Between $p_{tan}$ and $(\widetilde{p}_{-2})^{\IR}$}

The next step consists in expressing the  projector $(\widetilde{p}_{-2})^{\IR}$ in
(\ref{tilpro-2carrea}) as a sum of three pieces, just as we have done for the tangent
projector in (\ref{protansum}). It turns out that
\be\label{tilpro-2carreasum}
(\widetilde{p}_{-2})^{\IR} = \ket{W_1}\bra{W_1} + \ket{W_2}\bra{W_2} +
\ket{W_3}\bra{W_3}~,
\ee
with
\bea\label{components}
&& \bra{W_1} = {1 \over \sqrt{2}}
\left(~1 - (x_1)^2~;~~ 0~;~~ -x_3~;~~ x_1 x_2~;~~ - x_1 x_3~;~~ x_2 \right)~, 
\nn \\  
&& \bra{W_2} = {1 \over \sqrt{2}}
\left(~-x_1 x_2~;~~ x_3~;~~ 0~;~~ -1 + (x_2)^2~;~~ -x_2 x_3~;~~ -x_1 \right)~, 
   \nn \\ 
&& \bra{W_3} = {1 \over \sqrt{2}}
\left(~-x_1 x_3~;~~ -x_2~;~~ x_1~;~~ x_2 x_3~;~~ 1 - (x_3)^2~;~~ 0 \right)~. 
\eea
These three vector-valued functions are not independent since,
\be
\sum_{l=0}^3 ~x_l \bra{W_l} = 0~.
\ee
There is a simple relation between the three vector-valued functions $W_l$ in
(\ref{components}) and the corresponding $V_l$ in (\ref{infrot}). Indeed, 
\be
u \ket{V_l} = \ket{W_l}~, ~~l=1,2,3~,
\ee
with the matrix-valued function $u$ given by
\be\label{pariso} 
u = {1 \over \sqrt{2}}
\left(
\begin{array}{ccc}
0                 & -x_3      & x_2          \\         
~&~&~         \\
1 - (x_1)^2 & -x_1 x_2  & - x_1 x_3      \\ 
~&~&~         \\
- x_1 x_2     & 1 - (x_2)^2  & -x_2 x_3  \\ 
~&~&~         \\
-x_3          & 0        &   x_1         \\
~&~&~        \\
-x_2 & x_1 & 0        \\
~&~&~          \\
-x_1 x_3 & -x_2 x_3  & 1 - (x_3)^2   \\ 
\end{array}
\right)~.
\ee

The matrix $u$ turns out to be the partial isometry we are looking for. A lengthy
computation shows that
\be 
 u^\dagger u = p_{tan}~, ~~~~~u u^\dagger = (\widetilde{p}_{-2})^{\IR}~.
\ee 
This proves the equivalence between the two projectors $p_{tan}$ and
$(\widetilde{p}_{-2})^{\IR}$ and finishes the $K$-theory version of the 
isomorphism between the tangent bundle $TS^2$ and the real form of the
complex charge $2$ monopole bundle over $S^2$.

\bigskip\bigskip\bigskip
\bigskip\bigskip\bigskip
\noindent
{\bf Acknowledgments}. This work was motivated by conversations with P. Hajac.
I am grateful to him for several useful discussions and suggestions.

\vfill\eject
\bibliographystyle{unsrt}

\end{document}